\documentclass[letterpaper,12pt]{article}
\usepackage{lscape}
\usepackage{amsmath}
\usepackage{amssymb}
\usepackage[dvips]{graphicx,epsfig}
\usepackage{psfrag}
\newcommand{\im}{{\mathbb{I}}{\mathrm{m}}}
\newcommand{\re}{{\mathbb{R}}{\mathrm{e}}}

\title{Electromagnetic quasinormal modes of $D$-dimensional black holes}

\author{A.\ L\'opez-Ortega\thanks{Electronic address: alfredo@fis.cinvestav.mx} \\Departamento de F\'{\i}sica, CINVESTAV IPN\\ Apdo. Postal 14-740, 07000 M\'exico D. F., M\'exico.  }

\begin{document}

\maketitle

\begin{abstract}

Using the monodromy method we calculate the asymptotic quasinormal (QN) frequencies of an electromagnetic field moving in $D$-dimensional Schwarzschild and Schwarzschild de Sitter (SdS) black holes ($D\geq 4$). For the $D$-di\-men\-sion\-al Schwarzschild anti-de Sitter (SadS) black hole we also compute these frequencies with a similar method. Moreover, we calculate the electromagnetic normal modes of the $D$-dimensional anti-de Sitter (AdS) spacetime.

\end{abstract}



\section{Introduction}
\label{Section 1}

The dynamics of classical fields in higher dimensional black holes has been recently investigated, mainly motivated by the Brane World scenario in String Theory and the study of the higher dimensional features of General Relativity (see Refs.\ \cite{Frolov:2002xf}-\cite{Lopez-Ortega:2006} for some examples). These papers have studied the separability properties of the wave equations in higher dimensional backgrounds \cite{Frolov:2002xf}-\cite{Lopez-Ortega:2003gi}, the stability against small perturbations of several black holes \cite{Gibbons:2002pq}-\cite{Gleiser:2005ra}, their greybody factors \cite{Cardoso:2005vb}-\cite{Creek:2006ia}, and quasinormal modes (QNMs) \cite{Motl:2003cd}-\cite{Lopez-Ortega:2006}.

The QNMs of a black hole are solutions to the wave equations that satisfy physical boundary conditions near and far from the black hole horizon (for more details see reviews \cite{Kokkotas:1999bd}). For an asymptotically flat black hole they fulfill: (i) the field is purely ingoing near the horizon; (ii) the field is purely outgoing at infinity. For asymptotically not flat spacetimes, the boundary condition imposed at infinity can be different \cite{Birmingham:2001pj}, \cite{Dasgupta:1998jg}.

The numerical simulations provide evidence that no matter how we perturb a black hole its response will be dominated by the QNMs in some stages; this fact motivated the analysis of them due to its possible astrophysical importance \cite{Kokkotas:1999bd}. Furthermore, it was suggested that the QNMs can be a useful tool to study some aspects of the AdS-CFT and dS-CFT correspondences \cite{Horowitz:1999jd}-\cite{Abdalla:2002rm}, and even quantum gravity \cite{Hod:1998vk}, \cite{Dreyer:2002vy}, \cite{Birmingham:2003wa}.

As is well known, there are spacetimes in which an exact calculation of the quasinormal (QN) frequencies is possible \cite{Natario:2004jd}-\cite{Molina:2003ff}, \cite{Lopez-Ortega:2006}, \cite{Birmingham:2001pj}, \cite{Abdalla:2002hg}, \cite{Abdalla:2002rm}, \cite{Brady:1999wd}-\cite{Cardoso:2001hn}, but in many physically relevant backgrounds such exact calculation is not possible. Therefore there are many methods to calculate its values \cite{Kokkotas:1999bd}, for example: (i) the WKB approximation \cite{Schutz:1985yo}, \cite{Iyer:1986np}, \cite{Iyer:1986nq}, \cite{Konoplya:2003ii}; (ii) the phase integral method \cite{Andersson:1992yo}-\cite{Andersson:2003fh}; (iii) to substitute the effective potential which appears in the Schr\"o\-ding\-er type equation with an exactly solvable potential \cite{Mashhoon:1984yo}; (iv) the use of continued fractions \cite{Leaver:1985ax}-\cite{Motl:2002hd}; (v) the monodromy method \cite{Motl:2003cd} (see below).

Even if they are not astrophysically relevant, the recent computations of the asymptotic QN frequencies were motivated by Hod's conjecture \cite{Hod:1998vk}, which proposes that their real parts are related to the fundamental quanta of mass and angular momentum. An essential ingredient of Hod's conjecture is that the real part of the asymptotic QN frequencies must be proportional to the logarithm of an integer. Moreover, in Ref.\ \cite{Dreyer:2002vy} Dreyer proposed that this quantity is also related to the Barbero-Immirzi parameter of quantum gravity. Therefore, the asymptotic QN frequencies for several black holes have been computed in Refs.\ \cite{Motl:2003cd}, \cite{Birmingham:2003rf}, \cite{Natario:2004jd}, \cite{Ghosh:2005aq}, \cite{Cardoso:2003vt}, \cite{Cho:2005yc}, \cite{Andersson:2003fh}, \cite{Nollert:1993aa}, \cite{Motl:2002hd}, \cite{Cardoso:2004up}--\cite{Musiri:2003bv}; in these references different methods have been applied.\footnote{A criticism of these proposals is presented in Refs.\ \cite{Domagala:2004jt}.}

In the monodromy technique we analytically continue the radial coordinate into the complex plane and use information about the potentials and the tortoise coordinate around the origin, the horizons, and the infinity \cite{Motl:2003cd}. This method allows us to compute the asymptotic value of the QN frequencies for several black holes \cite{Motl:2003cd}, \cite{Birmingham:2003rf}, \cite{Natario:2004jd}, \cite{Ghosh:2005aq}, \cite{Cardoso:2003vt}.

For $D$-dimensional spacetimes ($D\geq4$), the monodromy method has been used. Exploiting it Motl in \cite{Motl:2003cd} computes the asymptotic QN frequencies of a scalar field in Schwarzschild background. The corresponding frequencies for the three types of gravitational perturbations moving in the same spacetime were calculated by Birmingham in \cite{Birmingham:2003rf}. The extension of these results was presented in \cite{Natario:2004jd} by Nat\'ario and Schiappa, who compute the asymptotic QN frequencies of the gravitational perturbations of the Schwarzschild, SdS and SadS black holes. In Refs.\ \cite{Ghosh:2005aq}, \cite{Das:2004db}, applying the monodromy technique, these frequencies of a massless Klein Gordon field (and therefore of a tensor gravitational perturbation) moving in asymptotically flat, de Sitter, and anti-de Sitter spherically symmetric black holes were calculated. It is convenient to notice that in \cite{Ghosh:2005aq} the frequencies of asymptotically flat and de Sitter spacetimes were found simultaneously.

In the extensive paper \cite{Natario:2004jd} the asymptotic QN frequencies for an electromagnetic field moving in $D$-dimensional Schwarzschild, Schwarzschild de Sitter (SdS), and Schwarzschild anti-de Sitter (SadS) black holes were not calculated. As already noted in Refs.\ \cite{Motl:2003cd}, \cite{Natario:2004jd}, if in the result for the asymptotic QN frequencies of the Reissner-Nordstr\"on background we take limit $Q \to 0$, where $Q$ stands for the electric charge of the black hole, we do not get that for the gravitational perturbations of the Schwarzschild spacetime.\footnote{The explication for this fact was proposed in \cite{Andersson:2003fh}. Recently a explicit calculation confirmed this proposal  (third reference in \cite{Cho:2005yc}), and we refer the reader to these works, where a more detailed and careful discussion can be found on the issues involved when the limit $Q \to 0$ is taken.} We believe that this is also true for the electromagnetic field. Moreover, the electromagnetic field has features different from the Klein-Gordon field and gravitational perturbations, which makes it worth studying. Therefore it is convenient to calculate separately the value of its asymptotic QN frequencies.

Here, following closely Refs.\ \cite{Motl:2003cd}, \cite{Birmingham:2003rf}, \cite{Natario:2004jd}, we exploit the monodromy technique to compute the asymptotic QN frequencies of an electromagnetic field propagating in $D$-dimensional Schwarzschild and SdS black holes and with a similar method we also calculate these frequencies for the $D$-dimensional SadS black hole. In this paper, we include the Schwarzschild black hole for completeness, since in several references this case is already studied (see for example Ref.\ \cite{Cardoso:2003vt}, \cite{Andersson:2003fh}, \cite{Musiri:2005ev}).

We also compute the normal frequencies of an electromagnetic field moving in $D$-dimensional anti-de Sitter (AdS) spacetime. They have been previously calculated in Refs.\ \cite{Natario:2004jd}, \cite{Cardoso:2003cj}, \cite{Breitenlohner:1982bm}-\cite{Avis:1977yn} for other fields, in four and higher dimensions.

A brief outline of this paper is the following. In Section \ref{Section 2} we concisely review the equations governing the dynamics of an electromagnetic field propagating in a higher dimensional spherically symmetric spacetime. In Sections \ref{Section 3}, \ref{Section 4}, and \ref{Section 5} we successively calculate the values of its asymptotic QN frequencies moving in $D$-dimensional Schwarzschild, SdS, and SadS black holes. In Section \ref{Section 6} we calculate the normal frequencies of the same field propagating in $D$-dimensional AdS spacetime and analyze some related facts. Finally, in Section \ref{Section 7} we discuss the main results.

\section{Electromagnetic field}
\label{Section 2}

By using a Feynman gauge, Crispino, \textit{et al.\ }in Ref.\ \cite{Crispino:2000jx} found that the equations of motion for an electromagnetic field propagating in a $D$-dimensional spherically symmetric and static spacetime can be simplified to a decoupled set of radial differential equations. Here we briefly review their main results.\footnote{For other studies of the electromagnetic field dynamics in higher dimensional spacetimes see Refs.\ \cite{Kodama:2003kk}, \cite{Ishibashi:2004wx}, \cite{Guven hertz potential}.}

In Schwarzschild's coordinates the metric of a $D$-dimensional spherically symmetric and static spacetime takes the form 
\begin{equation} \label{e:metric-general}
{\rm d}s^2= f(r)\, {\rm d}t^2 - h(r)\,{\rm d} r^2 -r^2 {\rm d} \Omega^2_{D-2},
\end{equation}
where ${\rm d} \Omega^2_{D-2}$ stands for the metric of the $(D-2)$-dimensional unit sphere. In the following sections we shall study the electromagnetic QN frequencies of the $D$-di\-men\-sion\-al Schwarzschild, SdS, SadS black holes, and the normal frequencies of the anti-de Sitter spacetime. Thus we can take
\begin{equation} \label{e:f equal to h}
h(r) = \frac{1}{f(r)},
\end{equation} 
and then the function $f(r)$ is equal to
\begin{eqnarray}
f(r) &= \left\{ \begin{array}{l} 1 - \frac{2M}{r^{D-3}} ,\quad \qquad \qquad \textrm{Schwarzschild black hole}, \\ 1 - \frac{2M}{r^{D-3}} - \lambda r^2 ,\qquad \,\,\textrm{SdS black hole}, \\ 1 - \frac{2M}{r^{D-3}} + \lambda r^2, \,\,\qquad \textrm{SadS black hole}, \\ 1  + \lambda r^2, \,\,\,\quad\qquad \qquad \textrm{AdS spacetime}, \end{array} \right.
\end{eqnarray} 
where $\lambda > 0$ ($M$) is a constant related to the cosmological constant (black hole mass).

In Ref.\ \cite{Crispino:2000jx} Crispino \textit{et al.\ }showed that the so-called modes \textbf{I} and \textbf{II} include all the physical degrees of freedom of an electromagnetic field moving in a $D$-dimensional spherically symmetric spacetime with metric (\ref{e:metric-general}).\footnote{In the notation of Ref.\ \cite{Kodama:2003kk}, the physical modes \textbf{I} and \textbf{II} correspond to scalar type and vector type electromagnetic perturbations respectively.} For the modes \textbf{I} the vector potential is given by Eq.\ (2.17) of \cite{Crispino:2000jx} and if the condition (\ref{e:f equal to h}) holds, then the radial function $R_{1}$ is a solution to the differential equation
\begin{eqnarray} \label{e:radial-I}
\frac{1}{r^2}\frac{{\rm d}}{{\rm d} r}\left[f(r) r^{4-D} \frac{{\rm d}}{{\rm d} r}\left( r^{D-2} R_{1} \right) \right] + \left[\frac{\omega^2}{f(r)}-\frac{l(l+D-3)}{r^2} \right] R_{1} = 0.
\end{eqnarray} 
On the other hand, for the modes \textbf{II} the vector potential is given by Eq.\ (2.20) of \cite{Crispino:2000jx} and when the condition (\ref{e:f equal to h}) holds, the radial function $R_{2}$ is a solution to the differential equation
\begin{equation} \label{e:radial-II}
\frac{1}{r^{D-4}} \frac{{\rm d}}{{\rm d} r} \left[f(r) r^{D-4} \frac{{\rm d}}{{\rm d} r} R_{2} \right]  + \left[ \frac{\omega^2}{f(r)}- \frac{(l+1)(l+D-4)}{r^2}\right] R_{2} = 0 .
\end{equation} 

In Ref.\ \cite{Crispino:2000jx} was also shown that Eqs.\ (\ref{e:radial-I}) and (\ref{e:radial-II}) can be transformed to the Schr\"o\-ding\-er type equations
\begin{equation} \label{e:Schrodinger-equation-I-II}
\left\{ \frac{{\rm d}^2 }{{\rm d} x^2}  + \omega^2 - V_{1,2}(r)\right\} \Phi_{1,2}(x) = 0,
\end{equation} 
where $x$ stands for the tortoise coordinate
\begin{equation} \label{e:tortoise-coordinate}
x = \int \frac{{\rm d} r}{f(r)},
\end{equation} 
$\Phi_{1}(x)$ ($\Phi_{2}(x)$) is related to $R_{1}$ ($R_{2}$) by
\begin{align}\label{e: R1 to Phi}
R_{1} = \frac{\Phi_{1}(x)}{r^{D/2}},  \qquad \qquad  
R_{2}= \frac{\Phi_{2}(x)}{r^{(D-4)/2}}, 
\end{align} 
and the effective potentials are equal to\footnote{In the following we write $f(r)$ simply as  $f$. }
\begin{align} \label{e:effective-potentials-I-II}
V_{1}(r) &= \frac{l(l+D-3)f}{r^2} + \frac{(D-2)(D-4)}{4r^2}f^2 - \frac{(D-4)f}{2r}\frac{{\rm d}f}{{\rm d}r} , \\
V_{2}(r) &= \frac{(l+1)(l+D-4)f}{r^2} + \frac{(D-4)(D-6)}{4r^2}f^2 + \frac{(D-4)f}{2r}\frac{{\rm d}f}{{\rm d}r}, \nonumber
\end{align} 
when the condition (\ref{e:f equal to h}) holds.

In Sections \ref{Section 3}, \ref{Section 4}, and \ref{Section 5} we shall need the behavior of the effective potentials (\ref{e:effective-potentials-I-II}) near $r=0$ for Schwarzschild, SdS, and SadS spacetimes. In the three cases we can show that in this region these potentials take the form \cite{Motl:2003cd}, \cite{Birmingham:2003rf}, \cite{Natario:2004jd}, \cite{Cardoso:2003vt}
\begin{equation} \label{l: potential near zero}
V_1(x) \sim \frac{j_1^2-1}{4 \,x^2},
\end{equation} 
(and a corresponding expression for $V_2(x)$). In Eq.\ (\ref{l: potential near zero}) $x$ stands for the approximate value of the tortoise coordinate for the Schwarzschild, SdS, and SadS black holes  near $r=0$,
\begin{equation} \label{l: torotise near origin}
x \sim - \frac{r^{D-2}}{2(D-2) M},
\end{equation} 
and the parameters $j_1$ and $j_2$ are equal to
\begin{align}
j_1 = \frac{2(D-3)}{D-2},  \qquad \qquad j_2 = \frac{2}{D-2}.
\end{align} 
Notice that the relation $j_1 + j_2 = 2$ holds.  

The values of $j_1$ and $j_2$ are different from those obtained for the tensor and scalar gravitational perturbations ($j=0$) and vector gravitational perturbations ($j=2$) moving in the same backgrounds \cite{Motl:2003cd}, \cite{Birmingham:2003rf}, \cite{Natario:2004jd}. We also notice that $j_1 \to 2 $ and $j_2 \to 0$ as $D \to \infty$. Thus, in this limit, we expect that the results for the asymptotic QN frequencies of the electromagnetic perturbations are almost indistinguishable from those found for the gravitational perturbations propagating in the same background. Furthermore, both $j_1$ and $j_2$ are equal to 1 when $D=4$.

It is convenient to note that the solutions to the Schr\"o\-ding\-er type equation (\ref{e:Schrodinger-equation-I-II}) with potentials (\ref{l: potential near zero}) take the form \cite{b:DE-books}
\begin{equation} \label{e: aproximate solution 0}
\Phi_1 (x) \sim B_+ \sqrt{2 \pi \omega x}\,\, J_{\tfrac{j_1}{2}}(\omega x) + B_- \sqrt{2 \pi \omega x} \,\,J_{- \tfrac{j_1}{2}}(\omega x) , 
\end{equation}
where $B_+$ and $B_-$ are constants and $J_\nu(z)$ denotes the Bessel function of order $\nu$, (there is also a corresponding solution for $\Phi_2 (x)$).

As is well known, near zero the Bessel functions satisfy
\begin{equation} \label{e: Bessel 2}
J_\nu(z) \sim z^\nu w(z),
\end{equation} 
where $w(z)$ is an even holomorphic function. For $z \gg 1$ and $z \ll -1$ these special functions also hold 
\begin{equation} \label{e: Bessel 1}
J_\nu (z) \sim \sqrt{\frac{2}{\pi z}} \cos \left( z - \frac{\nu \pi}{2} - \frac{\pi}{4} \right), \qquad \quad z \gg 1,
\end{equation} 
\begin{equation} \label{e: Bessel 3}
J_\nu (z) \sim \sqrt{\frac{2}{\pi z}} \cos \left( z + \frac{\nu \pi}{2} + \frac{\pi}{4} \right), \qquad \quad z \ll - 1.
\end{equation} 
Thus for $x \gg 1$ we can write the function $\Phi(x)$ of (\ref{e: aproximate solution 0}) in the form
\begin{equation} \label{l: approximate solution II}
\Phi_1 (x)\sim \left(B_+ \textrm{e}^{-i \alpha_+} + B_- \textrm{e}^{-i \alpha_-}\right)\textrm{e}^{i \omega x} + \left(B_+ \textrm{e}^{i \alpha_+} + B_- \textrm{e}^{i \alpha_-}\right)\textrm{e}^{- i \omega x},
\end{equation} 
where the phase shifts $\alpha_{\pm}$ are equal to
\begin{equation}
\alpha_{\pm} = \frac{\pi}{4}(1 \pm j_1).
\end{equation}

\section{$D$-dimensional Schwarzschild black hole}
\label{Section 3}

Our aim in this section is to compute by using the monodromy technique the asymptotic QN frequencies of an electromagnetic field moving in $D$-dimensional Schwarzschild background. At this point it is convenient to mention that our result for the asymptotic QN frequencies was already given in Refs.\ \cite{Motl:2003cd}, \cite{Birmingham:2003rf}, \cite{Cardoso:2003vt} (see also \cite{Andersson:2003fh}). Here we present it for completeness. The numerical evidence shows that for this spacetime the real and imaginary parts of the asymptotic QN frequencies hold $\im (\omega) \gg \re (\omega)$ \cite{Cardoso:2003vt}, \cite{Leaver:1985ax}, \cite{Nollert:1993aa}.

We closely follow Refs.\ \cite{Motl:2003cd}, \cite{Birmingham:2003rf}, \cite{Natario:2004jd} to perform the calculation and here we omit some technical details. Let us find the asymptotic QN frequencies with $\re (\omega) > 0$ for the modes \textbf{I}; thus we take the contour in the complex $r$-plane drawn in Figure 1 of Ref.\ \cite{Natario:2004jd} (see also \cite{Motl:2003cd}).\footnote{A numerical calculation of the Stokes lines corresponding to the Schwarzschild black hole in dimensions $D=4,5,6,7,$ is presented in Figure 2 of Ref.\ \cite{Natario:2004jd}.} We only remark that the contour used in the monodromy calculation depends upon the spacetime and is almost independent of the spacetime dimension (nevertheless see Section \ref{Section 4} below).

For the Schwarzschild QNMs, near infinity the field must be purely outgoing. Thus, from Eq.\ (\ref{l: approximate solution II}), the quantities $B_\pm$ must satisfy
\begin{equation} \label{l: condition 1 for B S}
\left(B_+ \textrm{e}^{-i \alpha_+} + B_- \textrm{e}^{-i \alpha_-}\right) = 0.
\end{equation} 
Moreover, the monodromy of the function $\textrm{e}^{- i \omega x}$ around the contour of Figure 1 in \cite{Natario:2004jd} is equal to 
\begin{equation} \label{l: condition 2 for B S}
\frac{B_+ \textrm{e}^{i 5 \alpha_+} + B_- \textrm{e}^{i 5 \alpha_-}}{B_+ \textrm{e}^{i \alpha_+} + B_- \textrm{e}^{i \alpha_-}} \textrm{e}^{-\pi \omega/ \kappa},
\end{equation} 
while far from the horizon the function is $\textrm{e}^{i \omega x}$ is exponentially small and its coefficient cannot be trusted in our approximation.

Notice that we can deform this contour to a small circle around the black hole horizon without crossing any singularities. Imposing the boundary conditions, we get that the clockwise monodromy of the function $\Phi_1(x)$ around of the new contour is equal to $\textrm{e}^{\pi \omega/ \kappa}$. Thus, taking into account this result and Eqs.\ (\ref{l: condition 1 for B S}) and (\ref{l: condition 2 for B S}), the asymptotic QN frequencies of the modes \textbf{I} are equal to \cite{Motl:2003cd}, \cite{Birmingham:2003rf}
\begin{equation} \label{l: asymptotic QN S}
\frac{\omega}{\kappa} = \frac{1}{2 \pi} \ln (1 + 2 \cos (\pi j_1)) + i\left(n + \frac{1}{2} \right), 
\end{equation} 
where $n=0,1,2,\dots$.

This equation was given by Motl in \cite{Motl:2003cd} and by Birmingham in \cite{Birmingham:2003rf}, but they only applied this formula to fields with integer values of $j$ and do not note that it applies to an electromagnetic field moving in $D$-dimensional Schwarzschild black hole. In Ref.\ \cite{Cardoso:2003vt}, Cardoso, \textit{et al.\ }used the method of Refs.\ \cite{Musiri:2003bv} to find Eq.\ (\ref{l: asymptotic QN S}) and its first order corrections; they also note that it applies to the electromagnetic field. The method of Refs.\ \cite{Musiri:2003bv} is an extension from that of \cite{Motl:2003cd}.

For the modes \textbf{II}, we known that $j_2 = 2 - j_1$, and therefore their asymptotic QN frequencies are also given by (\ref{l: asymptotic QN S}) \cite{Motl:2003cd}, \cite{Natario:2004jd}. As explained in the previous references, we can obtain the asymptotic QN frequencies whose real part is negative taking the contour in the opposite direction. 

In four dimensions $j_1 =j_2 = 1$ and so $\re (\omega) = 0$ as already calculated by Motl \cite{Motl:2003cd}. This result was verified numerically in Ref.\ \cite{Cardoso:2003vt}; there was shown that in this background the real part of the asymptotic QN frequencies for the electromagnetic field goes to zero as $n \to \infty$.

In the 5-dimensional Schwarzschild black hole for the real part of the asymptotic QN frequencies we obtain an ill-defined expression ($\ln(0)$) \cite{Cardoso:2003vt}. The cause of this result is probably that the real part of the QN frequencies rapidly goes to zero, as shown numerically in Ref.\ \cite{Cardoso:2003vt}, and asymptotically the condition $\re (\omega) > 0$ is not satisfied. We believe that this case must be studied in more detail.

Another interesting example is the 6-dimensional Schwarzschild spacetime. For this black hole $j_1=\tfrac{3}{2}$, $j_2=\tfrac{1}{2}$, and thus $\cos(\pi j_1) = \cos(\pi j_2) = 0$. Hence the expression (\ref{l: asymptotic QN S}) simplifies to
\begin{equation} \label{e: S QN frequencies 6D}
\frac{\omega}{\kappa} = i \left(n + \tfrac{1}{2}\right).
\end{equation} 
Thus the asymptotic QN frequencies of the electromagnetic field moving in the 6-di\-men\-sion\-al Schwarzschild black hole are purely imaginary as in four dimensions, (for other spacetimes whose QN frequencies are purely imaginary see Ref.\ \cite{Natario:2004jd}, \cite{Lopez-Ortega:2005ep}, \cite{Lopez-Ortega:2006ig}). 

It is convenient to remark that for an electromagnetic field moving in $D$-dimensional Schwarzschild black hole the real part of the asymptotic QN frequencies (\ref{l: asymptotic QN S}) is only proportional to the logarithm of a positive integer when $D = 4$, and $D= 6$ ($\ln(1)=0$).

\section{$D$-dimensional Schwarzschild de Sitter black hole}
\label{Section 4}

The SdS QNMs are solutions to the equations of motion that satisfy: (i) the modes are purely ingoing near the black hole horizon ($r_H$); (ii) the modes are purely outgoing near the cosmological horizon ($r_C$). The purpose of this section is to calculate the asymptotic QN  frequencies of an electromagnetic field propagating in this background. We first study the modes \textbf{I} following closely Ref.\ \cite{Natario:2004jd} and then we discuss the modes \textbf{II}. The numerical results obtained in Refs.\ \cite{Yoshida:2003zz}, \cite{Konoplya:2004uk} for the QN frequencies of the SdS black holes indicate that in the asymptotic limit the condition $\im (\omega) \gg \re (\omega)$ with $\im (\omega) \to \infty$ holds. 

We employ the same contour in the complex $r$-plane that in Ref.\ \cite{Natario:2004jd} to perform the monodromy calculation. It appears in Figure 7 of \cite{Natario:2004jd} and coincides with a Stokes line for $D\geq 6$. We only remark that the configuration of the Stokes lines for the $D$-dimensional SdS black hole near the origin ($r=0$) is equal to that of the Schwarzschild spacetime of the same dimension \cite{Natario:2004jd}. Nat\'ario and Schiappa in Ref.\ \cite{Natario:2004jd} also show that for $D\geq 6$ there is a Stokes line that cuts the real axis at $r_C > r > r_H$ and another that cuts the same axis at $r>r_C$ \cite{Natario:2004jd}.\footnote{The numerically calculated Stokes lines corresponding to the $D$-dimensional SdS black hole for $D=4,5,6,7,$ are found in Figure 8 of \cite{Natario:2004jd}.} In the rest of this section $\kappa_H$ and $\kappa_C$ stand for the surface gravity of the black hole horizon and the cosmological horizon respectively, ($\kappa_C<0$).

As already noted in Section \ref{Section 2} (see also Refs.\ \cite{Natario:2004jd}, \cite{Cardoso:2004up}) the effective potential and the tortoise coordinate for the SdS black hole near the origin take the forms (\ref{l: potential near zero}) and (\ref{l: torotise near origin}) respectively. Hence from Eqs.\ (\ref{e: Bessel 2}), (\ref{e: Bessel 1}), and (\ref{e: Bessel 3}) the function $\Phi_1(x)$ at point $\textbf{A}$ takes the form (\ref{l: approximate solution II}) and at point $\textbf{B}$ it is given by
\begin{equation} \label{e: Phi at point B}  
\Phi_1 (x) \sim \left(B_+ \textrm{e}^{i 7 \alpha_+} + B_- \textrm{e}^{i 7 \alpha_-}\right)\textrm{e}^{i \omega x} + \left(B_+ \textrm{e}^{i 5 \alpha_+} + B_- \textrm{e}^{i 5 \alpha_-}\right)\textrm{e}^{- i \omega x}.
\end{equation} 

We can show that the monodromy of the first term in the right hand side of Eq.\ (\ref{l: approximate solution II}) around the contour of Figure 7 in \cite{Natario:2004jd} is equal to
\begin{equation}
\frac{B_+ \textrm{e}^{i 7 \alpha_+} + B_- \textrm{e}^{i 7 \alpha_-}}{B_+ \textrm{e}^{- i \alpha_+} + B_- \textrm{e}^{-i \alpha_-}} \textrm{e}^{\pi \omega / \kappa_H + \pi \omega / \kappa_C}.
\end{equation} 
A similar expression holds for the monodromy of the second term in the right hand side of Eq.\ (\ref{l: approximate solution II}).

The contour used here can be deformed to one that only encloses the black hole horizon and the cosmological horizon without crossing any singularities. From the behavior of the effective potentials near $r_H$ and $r_C$, and taking into account the boundary conditions of the SdS QNMs, we find that the monodromy of the function $\Phi_1(x)$ around the new contour is equal to $\textrm{e}^{ \pi \omega / \kappa_H - \pi \omega / \kappa_C}$.

The two terms of Eq.\ (\ref{l: approximate solution II}) have the required monodromy if we impose the conditions
\begin{align}
\frac{B_+ \textrm{e}^{i 7 \alpha_+} + B_- \textrm{e}^{i 7 \alpha_-}}{B_+ \textrm{e}^{- i \alpha_+} + B_- \textrm{e}^{-i \alpha_-}} \textrm{e}^{\pi \omega / \kappa_H + \pi \omega / \kappa_C} &= \textrm{e}^{ \pi \omega / \kappa_H - \pi \omega / \kappa_C} ,\nonumber \\
\frac{B_+ \textrm{e}^{i 5 \alpha_+} + B_- \textrm{e}^{i 5 \alpha_-}}{B_+ \textrm{e}^{ i \alpha_+} + B_- \textrm{e}^{i \alpha_-}} \textrm{e}^{- \pi \omega / \kappa_H - \pi \omega / \kappa_C} &= \textrm{e}^{ \pi \omega / \kappa_H - \pi \omega / \kappa_C}.
\end{align} 
From the previous equations, the asymptotic QN frequencies of the electromagnetic field are implicitly determined by
\begin{equation} \label{e: QNMs SdS 6}
\sin\left(\frac{3 \pi}{2} j_1 \right) \cosh\left(\frac{\pi \omega}{\kappa_C} + \frac{\pi \omega}{\kappa_H} \right) + \sin\left(\frac{ \pi}{2} j_1 \right) \cosh\left(\frac{\pi \omega}{\kappa_C} - \frac{\pi \omega}{\kappa_H} \right) = 0.
\end{equation} 

This result is different from that given by Nat\'ario and Schiappa in \cite{Natario:2004jd}, since only the expression for the cases $j=0$ and $j=2$ was presented in the previous reference. We can obtain its formula for the asymptotic QN frequencies differentiating Eq.\ (\ref{e: QNMs SdS 6}) with respect to $j_1$ and then evaluating at $j_1=0$.

From the numerically calculated Stokes diagrams which appear in Figure 8 of Ref.\ \cite{Natario:2004jd} for the 4-dimensional and 5-dimensional SdS black holes, we see that in these two cases the configuration of the Stokes lines is different from that we supposed, as already noticed in Ref.\ \cite{Natario:2004jd}. The asymptotic QN frequencies of an electromagnetic field moving in 4-di\-men\-sion\-al SdS black hole were previously calculated in Ref.\ \cite{Cardoso:2004up} using a contour in the complex $r$-plane different from that of Figure 7 in Ref.\ \cite{Natario:2004jd} (see Figure 1 of \cite{Cardoso:2004up}). Nevertheless, for this case we can simplify the problem to that already studied and obtain the same expression for the asymptotic QN frequencies (\ref{e: QNMs SdS 6}) (see \cite{Natario:2004jd} for more details).

\begin{figure}[t]
\begin{center}
\psfrag{A}{$A$}
\psfrag{B}{$B$}
\psfrag{rh}{$r_H$}
\psfrag{rc}{$r_C$}
\psfrag{Stokes line}{Stokes line}
\psfrag{Equivalent}{Equivalent}
\psfrag{r-plane}{$r$-plane}
\psfrag{contour}{contour}
\psfrag{}{}
\includegraphics[scale=1]{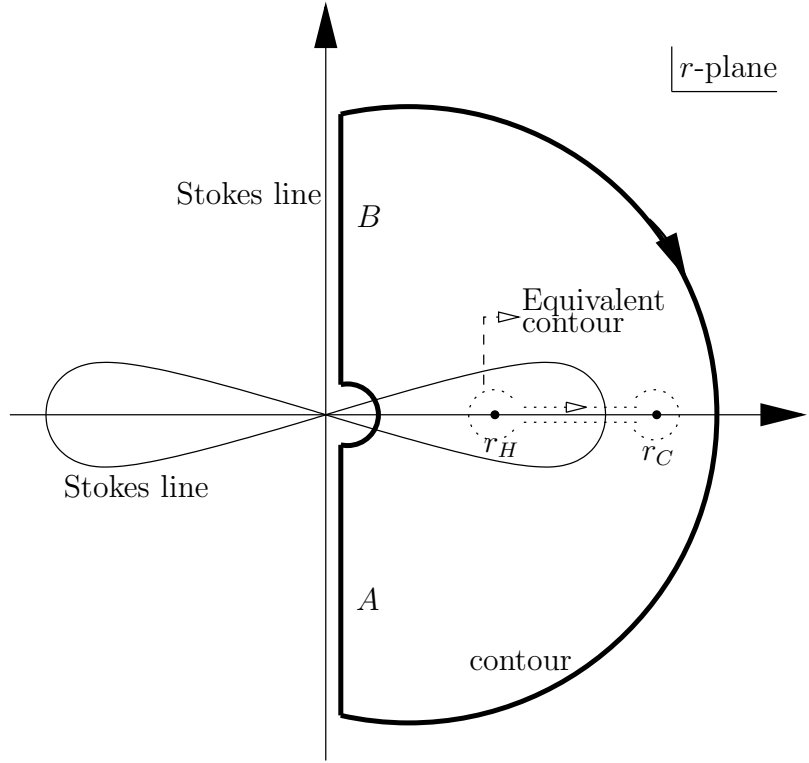}
\caption{Contour for the 5-dimensional Schwarzschild de Sitter black hole.}
\label{fig: three}
\end{center}
\end{figure}

Moreover, from Figure 8 of \cite{Natario:2004jd}, we note that for the 5-dimensional SdS black hole there are Stokes lines that go to infinity \cite{Natario:2004jd}. Therefore we must take a closed contour near infinity; it appears in Figure \ref{fig: three}. Near the singularity at $r=0$, in the branch containing the point $\textbf{A}$ the function $\Phi_1(x)$ takes the form (\ref{l: approximate solution II}) and at the branch holding the point $\textbf{B}$, $\Phi_1(x)$ is given by (\ref{e: Phi at point B}). According to Eqs.\ (\ref{e:tortoise-coordinate}) and (\ref{e:effective-potentials-I-II}) near infinity $x \sim x_0$, where $x_0$ is a complex constant and 
\begin{equation} \label{e: potential near infinity}
V_{1,2} (x) \sim \frac{j_{\infty}^2 -1 }{4(x-x_0)^2},
\end{equation} 
with $j_{\infty} = (D-5)$ for the modes \textbf{I} (equal to scalar gravitational perturbations) and $j_{\infty} = (D-3)$ for modes \textbf{II} (equal to vector gravitational perturbations) \cite{Natario:2004jd}. Hence, the function $\Phi_1(x)$ near infinity  can be written as in Eq.\ (\ref{e: aproximate solution 0}) if we make the changes $\omega x \to \omega (x-x_0)$, $B_\pm \to A_\pm$, and $j_1 \to j_\infty$.

Since we suppose that the frequency satisfies $\im (\omega)$ $\gg \re (\omega)$ with $\im(\omega)$ $\to \infty$, it is possible to show that near infinity $\omega (x-x_0) < 0$ at the branch holding $\textbf{A}$ and $\omega(x-x_0) > 0$ at the branch containing $\textbf{B}$. Therefore at this branch the function $\Phi_1(x)$ takes the form \cite{Natario:2004jd}
\begin{equation} \label{l: approximate solution infinity}
\Phi_1(x) \sim \left(A_+ \textrm{e}^{-i \beta_+} + A_- \textrm{e}^{-i \beta_-}\right)\textrm{e}^{i \omega (x-x_0)} + \left(A_+ \textrm{e}^{i \beta_+} + A_- \textrm{e}^{i \beta_-}\right)\textrm{e}^{- i \omega (x-x_0)},
\end{equation} 
where
\begin{equation}
\beta_{+} = \frac{\pi (1 + j_\infty)}{4}.
\end{equation} 

Now, taking the contour showed in Figure \ref{fig: three} beginning at $\textbf{A}$, we must note that near infinity if $r$ rotates clockwise an angle $\pi$, then the tortoise coordinate $x$ rotates anticlockwise by the same quantity. Exploiting these facts and taking into account the value of the parameter $j_\infty$ for the modes \textbf{I}, we find that in this branch,  the function $\Phi_1(x)$ becomes \cite{Natario:2004jd}
\begin{align} \label{l: approximate solution infinity rotate 2}
\Phi_1 (x) \sim - \left(A_+ \textrm{e}^{ - i  \beta_+} + A_- \textrm{e}^{- i  \beta_-}\right)& \textrm{e}^{i \omega (x-x_0)} \nonumber \\ 
&+ \left(A_+ \textrm{e}^{i \beta_+} + A_- \textrm{e}^{i \beta_-}\right)\textrm{e}^{- i \omega (x-x_0)}.
\end{align} 
Hence the coefficient of $\textrm{e}^{i \omega (x-x_0)}$ changes of sign, while the coefficient of $\textrm{e}^{- i \omega (x-x_0)}$ remains unchanged.\footnote{This also happens for the modes \textbf{II}.} 

As in the previous cases, we can deform the contour of Figure \ref{fig: three} to a small contour that enclose the black hole and cosmological horizons, without crossing any singularities. Thus the terms of Eq.\ (\ref{l: approximate solution II}) have the required monodromy if we impose the conditions \cite{Natario:2004jd}
\begin{eqnarray}
- \frac{B_+ \textrm{e}^{i 7 \alpha_+} + B_- \textrm{e}^{i 7 \alpha_-}}{B_+ \textrm{e}^{- i \alpha_+} + B_- \textrm{e}^{-i \alpha_-}}\, \textrm{e}^{\pi \omega / \kappa_H + \pi \omega / \kappa_C} = \textrm{e}^{ \pi \omega / \kappa_H - \pi \omega / \kappa_C}, \nonumber \\
\frac{B_+ \textrm{e}^{i 5 \alpha_+} + B_- \textrm{e}^{i 5 \alpha_-}}{B_+ \textrm{e}^{ i \alpha_+} + B_- \textrm{e}^{i \alpha_-}}\, \textrm{e}^{- \pi \omega / \kappa_H - \pi \omega / \kappa_C} = \textrm{e}^{ \pi \omega / \kappa_H - \pi \omega / \kappa_C}.
\end{eqnarray} 

From previous equations, the asymptotic QN frequencies of the electromagnetic field are implicitly determined by
\begin{equation} \label{e: QNMs SdS 5D} 
\sin\left(\frac{3 \pi}{2} j_1 \right) \sinh\left(\frac{\pi \omega}{\kappa_C} + \frac{\pi \omega}{\kappa_H} \right) - \sin\left(\frac{ \pi}{2} j_1 \right) \sinh\left(\frac{\pi \omega}{\kappa_H} - \frac{\pi \omega}{\kappa_C} \right) = 0.
\end{equation} 
This expression is different from that found in Ref.\ \cite{Natario:2004jd} since in this reference only the results for $j=0$ and $j=2$ were presented. We can obtain the finding of \cite{Natario:2004jd} differentiating Eq.\ (\ref{e: QNMs SdS 5D}) with respect to $j_1$ and then evaluating the derivative at $j_1=0$. For the modes \textbf{II} we find the same results for their asymptotic QN frequencies because $j_2 = 2 - j_1$ \cite{Motl:2003cd}, \cite{Birmingham:2003rf}, \cite{Natario:2004jd}. Also notice that if $\omega$ is a solution to Eqs.\ (\ref{e: QNMs SdS 6}) and (\ref{e: QNMs SdS 5D}), then $-\omega^*$ is also a solution. 

When $D=4$, we find $j_1 =j_2= 1$ and hence the asymptotic QN frequencies of the electromagnetic are equal to
\begin{equation} \label{e: QN frequency 4D SdS}
\omega = i n \kappa_H  \qquad \textrm{or} \qquad \omega = - i n \kappa_C,  \qquad n \in \mathbb{N}.
\end{equation} 
These values are equal to those\footnote{We note that in Ref.\ \cite{Cardoso:2004up} the second value for the SdS QN frequencies is $\omega = i n \kappa_C$, with $n \in \mathbb{N}$; thus $\im (\omega ) < 0$ since $\kappa_C$ is a negative number, but we must have $\im (\omega) > 0$ in order that the perturbations decay with time.} found in Ref.\ \cite{Cardoso:2004up} and we also point out that these results were numerically verified in Refs.\ \cite{Yoshida:2003zz}, \cite{Konoplya:2004uk}. For $D=6$, $j_1 = \tfrac{3}{2}$, $j_2=\tfrac{1}{2}$, and from Eq.\ (\ref{e: QNMs SdS 6}) we find that the asymptotic QN frequencies take the form
\begin{equation} \label{e:QN frequency 6D SdS}
\omega = i \kappa_H \left(n + \tfrac{1}{2} \right)  \qquad \textrm{or} \qquad \omega = - i \kappa_C \left(n + \tfrac{1}{2} \right),  \qquad n \in \mathbb{N}.
\end{equation} 
The asymptotic QN frequencies (\ref{e: QN frequency 4D SdS}) and (\ref{e:QN frequency 6D SdS}) are purely imaginary as those of the de Sitter spacetime \cite{Natario:2004jd}, \cite{Du:2004jt}, \cite{Lopez-Ortega:2006}, \cite{Lopez-Ortega:2006ig}.

To investigate the Schwarzschild limit of our results (\ref{e: QNMs SdS 6}) and (\ref{e: QNMs SdS 5D}) we note that in the same limit the following condition holds
\begin{equation} \label{e: Schwarzschild limit}
\textrm{e}^{ -\pi \omega / \kappa_C} \gg \textrm{e}^{ \pi \omega / \kappa_C},
\end{equation} 
since $\kappa_C \to 0^-$ \cite{Natario:2004jd}. Therefore, according to Eq.\ (\ref{e: Schwarzschild limit}), both Eqs.\ (\ref{e: QNMs SdS 6}) and (\ref{e: QNMs SdS 5D}) simplify to
\begin{equation}
\textrm{e}^{2 \pi \omega / \kappa_H} = - (1 + 2 \cos (\pi j_1)),
\end{equation} 
from which the  same expression (\ref{l: asymptotic QN S}) for the asymptotic QN frequencies of the Schwarzschild black hole can be found. So, for Eqs.\ (\ref{e: QNMs SdS 6}) and (\ref{e: QNMs SdS 5D}) this limit is well behaved as already noted in \cite{Natario:2004jd}, \cite{Cardoso:2004up} for the expressions calculated there.

On the other hand, to calculate the de Sitter limit of our results (\ref{e: QNMs SdS 6}) and (\ref{e: QNMs SdS 5D}) we note the following. In Ref.\ \cite{Natario:2004jd} is asserted that in de Sitter limit the surface gravity of the black hole horizon is given by
\begin{equation} \label{e:Natario kh}
\kappa_H \sim \frac{D-3}{2} r_H,
\end{equation} 
and then
\begin{equation}
\textrm{e}^{ \pi \omega / \kappa_H} \gg \textrm{e}^{ - \pi \omega / \kappa_H}.
\end{equation}
Nevertheless, it is well known that if the radius of a black hole goes to zero, then the surface gravity increases, which is different from the behavior found in Eq.\ (\ref{e:Natario kh}) for it. Therefore in the following we first calculate the de Sitter limit of $\kappa_H$ and $\kappa_C$ for the SdS black hole in order to calculate it for the asymptotic QN frequencies already found.

For calculating $\kappa_H$ when $2 M \ll 1$, we first note that in this limit $2M \sim r_H^{D-3}$ and then
\begin{equation}
\kappa_H =\frac{1}{2}f^{\prime}(r_H) \sim \frac{D-3}{2 r_H},
\end{equation} 
thus $\kappa_H \to \infty$ in the de Sitter limit. Moreover in the same limit \cite{Natario:2004jd}
\begin{equation}
\kappa_C \sim - \sqrt{\lambda},
\end{equation}   
since $r_C \sim \tfrac{1}{\sqrt{\lambda}} = L$ when $2 M \ll 1$.

Hence, in de Sitter limit $ \tfrac{\pi \omega}{\kappa_H} \sim 0$ holds (that is, $\textrm{e}^{\pi \omega / \kappa_H} \sim \textrm{e}^{- \pi \omega / \kappa_H} \sim 1$) and so for $j_1 \neq  q$, $q \in \mathbb{Z}$, we find that Eq.\ (\ref{e: QNMs SdS 6}) simplifies to
\begin{equation} \label{e: condition de Sitter limit}
\cos\left( \frac{i \pi \omega}{\kappa_C} \right) = 0,
\end{equation} 
which implies
\begin{equation} \label{e: de Sitter limit 6}
\tilde{\omega} = i \left( n + \tfrac{1}{2}\right), \qquad n \in \mathbb{N},
\end{equation} 
where $\tilde{\omega} = \omega L$. Taking the de Sitter limit of Eq.\ (\ref{e: QNMs SdS 5D}) with a similar procedure, we get the QN frequencies 
\begin{equation} \label{e: de Sitter limit 5}
\tilde{\omega} = i n, \qquad n \in \mathbb{N}.
\end{equation} 
For the 4-dimensional SdS black hole $j_1=1$, and so we cannot use Eq.\ (\ref{e: condition de Sitter limit}). For this case, taking the de Sitter limit we also find the result given in Eq.\ (\ref{e: de Sitter limit 5}).

In Ref.\ \cite{Natario:2004jd} is shown that its Eq.\ (3.28) determines the asymptotic QN frequencies of the gravitational perturbations moving in $D$-dimensional SdS black hole ($D \geq 6$). In de Sitter limit this equation simplifies to Eq.\ (\ref{e: condition de Sitter limit}) and so we obtain the QN frequencies (\ref{e: de Sitter limit 6}). Furthermore, the expression found in \cite{Natario:2004jd} for the asymptotic QN frequencies of the gravitational perturbations propagating in 5-dimensional SdS black hole gives the QN frequencies (\ref{e: de Sitter limit 5}) in the same limit.

From their result for the QN frequencies of a gravitational perturbation moving in the SdS black hole, Nat\'ario and Schiappa \cite{Natario:2004jd} calculated that the gap\footnote{According to Ref.\ \cite{Natario:2004jd} the offset is the real part of the emitted frequency and the gap is equal to the quantized increment in the inverse relaxation time.} is equal to $1/L$ in de Sitter limit. Moreover, they directly calculated its de Sitter QN frequencies in odd dimensions. From these last values they deduced that the gap is given by $2/L$. In Ref.\ \cite{Lopez-Ortega:2006} is shown that the de Sitter QN frequencies for a gravitational perturbation and an electromagnetic field in $D$ dimensions are purely imaginary (as already noticed in \cite{Natario:2004jd} for the gravitational perturbations in odd dimensions). Furthermore, in \cite{Lopez-Ortega:2006} is claimed that the gap is equal to $1/L$; but we can deduce from its Eqs.\ (19), (25), and (37) that in the limit $n \to \infty$ the gap is equal to $1/L$ for even $D$ while it is equal to $2/L$ for odd $D$. Hence, Eqs.\ (\ref{e: de Sitter limit 6}) and (\ref{e: de Sitter limit 5}) yield the correct gap only for even $D$.\footnote{We also notice that the de Sitter QN frequencies of the scalar gravitational perturbations and modes \textbf{I} are equal; this also happens for those of the vector gravitational perturbations and modes \textbf{II}. This fact was overlooked in Ref.\ \cite{Lopez-Ortega:2006}.}

Also, from Eqs.\ (\ref{e: de Sitter limit 6}) and (\ref{e: de Sitter limit 5}), the correct offset (zero) for the de Sitter QN frequencies is obtained. We also note that the de Sitter spacetime is not singular at $r=0$; hence the Stokes diagram for this spacetime must be different from those used in this section for the SdS background. Thus, for the asymptotic QN frequencies of the SdS black hole, these facts show that the de Sitter limit cannot be taken in a straightforward way as already noted in \cite{Natario:2004jd}, \cite{Lopez-Ortega:2006}.

\section{$D$-dimensional Schwarzschild anti-de Sitter black hole}
\label{Section 5}

For classical fields propagating in asymptotically anti-de Sitter spacetimes, the AdS-CFT correspondence recently has motivated the computation of their QNMs \cite{Horowitz:1999jd}, \cite{Cardoso:2001bb}, \cite{Siopsis:2004up}. The study of the electromagnetic perturbations in these spacetimes is relevant for the above mentioned correspondence since they can be seen as perturbations of some Supergravity gauge field. In this section, we calculate the asymptotic QN frequencies of an electromagnetic field moving in $D$-dimensional SadS black hole following closely Ref.\ \cite{Natario:2004jd}. We first study the modes \textbf{I} and then the modes \textbf{II}. 

Notice that in Refs.\ \cite{Musiri:2005ev}, \cite{Musiri:2003bv} the asymptotic QN frequencies for Schwarzschild anti-de Sitter black holes have been computed applying a similar technique. This method allow us to compute the first order corrections to the QN frequencies. Nevertheless, in \cite{Natario:2004jd}, \cite{Cardoso:2004up},  \cite{Musiri:2005ev}, \cite{Musiri:2003bv} the asymptotic QN frequencies (and its first order corrections) for an electromagnetic field moving in $D$-dimensional SadS black hole were not calculated when $D \geq 5$.

The SadS QNMs are solutions to the field equations that satisfy: (i) the modes are purely ingoing near the black hole horizon; (ii) the field vanish at infinity. The boundary conditions at infinity can be different \cite{Birmingham:2001pj}, \cite{Dasgupta:1998jg}, however we use (ii) because the energy of the perturbations is conserved \cite{Ghosh:2005aq}. The numerical results for the asymptotic QN frequencies of the SadS black hole show that these hold $\omega x_0 \in \mathbb{R}$ \cite{Natario:2004jd}, \cite{Cardoso:2004up}, \cite{Cardoso:2003cj}, \cite{Berti:2003ud}, and it is no longer true that $\im (\omega) \gg \re (\omega)$ as for the Schwarzschild and SdS black holes.

It is convenient to mention that the procedure applied in Ref.\ \cite{Natario:2004jd} (see also \cite{Cardoso:2004up}) is not a monodromy calculation since the contour used is not closed (see below). Nevertheless in \cite{Natario:2004jd}, \cite{Cardoso:2004up} the radial coordinate is analytically continued into the complex plane and they choose a contour from infinity to the black hole horizon along the Stokes lines. Thus, for the SadS spacetime, the solutions found to the Schr\"odinger type radial equations in different regions of the complex $r$-plane are matched along the Stokes lines in the appropriate limits.

To calculate the asymptotic QN frequencies of an electromagnetic field moving in SadS black hole we use the same the contour and conventions on the branch cuts that Nat\'ario and Schiappa in \cite{Natario:2004jd}. The contour appear in Figure 12 of this reference, (see also Ref.\ \cite{Cardoso:2004up}). Here we only remark that there is at least one Stokes line which goes to infinity and other that hit the $D-1$ horizons in spiraling form.\footnote{In Figures 13 and 14 of Ref.\ \cite{Natario:2004jd} are shown the numerically computed Stokes lines for the $D$-dimensional SadS black holes when $D=4,5,6,7$.}

For $r \to \infty$ the effective potentials take the form (\ref{e: potential near infinity}) with the same values of the parameter $j_\infty$ for the modes \textbf{I} and \textbf{II} which for the SdS black hole. This fact implies that we get the same solution near infinity. Thus, imposing the SadS QNMs boundary condition at infinity, we get that at the Stokes line containing the point $\textbf{B}$ the function $\Phi_1(x)$ near infinity takes the form
\begin{equation}
\Phi_1(x) \sim A_+ \sqrt{2 \pi \omega (x -x_0)} \,\, J_{\tfrac{j_\infty}{2}} (\omega (x- x_0)).
\end{equation} 
Taking into account that near infinity at this Stokes line $\omega (x-x_0) < 0$, we obtain that in the limit $|\omega(x-x_0) |\gg 0$ (thus, for smaller $|r|$) the function $\Phi_1(x)$ becomes
\begin{equation} \label{e: phi infinity small radial}
\Phi_1(x) \sim A_+  \textrm{e}^{i \beta_+ + i \omega (x-x_0)} + A_+  \textrm{e}^{-i \beta_+ - i \omega (x-x_0)}.
\end{equation} 

According to the conventions of Nat\'ario and Schiappa \cite{Natario:2004jd}, near $r=0$, at the branch containing the point $\textbf{B}$, $\omega x > 0$. In the same limit in which Eq.\ (\ref{e: phi infinity small radial}) holds ($\omega x$ sufficiently large), we also find that $\Phi_1(x)$ is given by Eq.\ (\ref{l: approximate solution II}). To match it with the function (\ref{e: phi infinity small radial}) the following condition must be fulfilled
\begin{equation} \label{e: condition 1 SdS QNMs}
\textrm{e}^{i \omega x_0 - i \beta_+}(B_+ \textrm{e}^{-i \alpha_+} + B_-\textrm{e}^{-i \alpha_-}) = \textrm{e}^{- i \omega x_0 + i \beta_+}(B_+ \textrm{e}^{i \alpha_+} + B_-\textrm{e}^{i \alpha_-}).
\end{equation} 

We must rotate from the Stokes line holding $\textbf{B}$ to that containing $\textbf{A}$ to satisfy the boundary condition at the black hole horizon. According to Eqs.\ (\ref{e: aproximate solution 0}) and (\ref{e: Bessel 2}) the function $\Phi_1(x)$ in this branch takes the form
\begin{equation} \label{e: expansion phi}
\Phi_1(x) \sim (B_+ \textrm{e}^{-i \alpha_+} + B_-\textrm{e}^{-i \alpha_-})\textrm{e}^{i \omega x} + (B_+ \textrm{e}^{-i 3 \alpha_+} + B_-\textrm{e}^{-i 3 \alpha_-})\textrm{e}^{- i \omega x}.
\end{equation} 
The previous solution can be propagated along this Stokes line up to the black hole horizon; where the boundary condition for the SadS QNMs can be imposed; as a consequence we get that the constants $B_+$ and $B_-$ must satisfy
\begin{equation} \label{e: condition 2 SdS QNMs}
B_+ \textrm{e}^{-i 3 \alpha_+} + B_-\textrm{e}^{-i 3 \alpha_-} = 0 .
\end{equation} 

When $j_1 \neq 2q + 1$, $q \in \mathbb{Z}$, from Eqs.\ (\ref{e: condition 1 SdS QNMs}) and (\ref{e: condition 2 SdS QNMs}), the asymptotic QN frequencies of the modes {\bf I} are determined by
\begin{equation}  \label{e: QN frequencies D SadS}
\omega x_0 = \frac{\pi}{4}(D + 1) + n\pi + \frac{1}{2i}\ln\left(2 \cos\frac{\pi j_1}{2} \right), \qquad n \in \mathbb{N}.
\end{equation} 

The cases of spacetime dimension $D=4$ and $D=5$ must be studied separately for the modes \textbf{I} \cite{Natario:2004jd}. When $D=4$ the parameter $j_\infty$ is equal to $-1$, but in our treatment was implicitly assumed that $j_\infty > 0$. We overcome this problem remarking that only the condition $j_\infty^2 =1$ must be satisfied; thus we can take $j_1 = 1$ \cite{Natario:2004jd}. For this case another problem is that $j_1 =1$ and then Eq.\ (\ref{e: QN frequencies D SadS}) cannot be used. Hence, as already noted in Ref.\ \cite{Musiri:2005ev}, to get a well defined result we must calculate the first correction to the QN frequencies, since in four dimensions for the effective potential of the modes \textbf{I} and \textbf{II} the approximation (\ref{l: potential near zero}) vanishes. As previously mentioned, in Ref.\ \cite{Musiri:2005ev}  is used an extension of the method presented in Ref.\ \cite{Motl:2003cd}. 

For the modes \textbf{I} in 5-dimensional SadS spacetime, $j_\infty = 0$. Using the procedure explained in \cite{Natario:2004jd}, we find that there is no constraint on the frequency; therefore the spectrum of its asymptotic QN frequencies is continuous, as for scalar type gravitational perturbations. 

Recalling that $j_2 = 2 - j_1$ and exploiting a similar method, we also find the expression (\ref{e: QN frequencies D SadS}) for the asymptotic QN frequencies of the modes \textbf{II}. Equation (\ref{e: QN frequencies D SadS}) implies that the asymptotic QN frequencies depend on the features of the SadS black hole studied due to the factor $x_0$. This equation is different from that obtained in \cite{Natario:2004jd} for the gravitational perturbations propagating in the SadS black hole since in previous reference only the result for $j=0$ and $j=2$ was presented; it is similar to that for the asymptotic QN frequencies of the Reissner-Nordstr\"om anti-de Sitter black hole also found there. Furthermore, Musiri, \textit{et al.\ }in Ref.\ \cite{Musiri:2005ev} obtained a similar expression in their study of the first order corrections for the QN frequencies of the gravitational perturbations moving in $D$-dimensional SadS background.

\section{$D$-dimensional anti-de Sitter spacetime}
\label{Section 6}

In this section, for an electromagnetic field propagating in the AdS background we calculate its normal modes; they are solutions to the equations of motion that satisfy the boundary conditions: (i) the modes are regular at $r=0$; (ii) they vanish as $r \to \infty$ \cite{Natario:2004jd}, \cite{Cardoso:2003cj}. In Refs.\ \cite{Breitenlohner:1982bm}, \cite{Burgess:1984ti} these modes were discussed due to its usefulness in analyzing the stability properties of ground states in Supergravity theories against small fluctuations which vanish sufficiently fast at spatial infinity. More recently the AdS-CFT correspondence also motivated their study \cite{Cardoso:2003cj}, \cite{Cotaescu:1998ts}.

We first notice that in AdS spacetime the effective potentials for the modes \textbf{I} and \textbf{II} of the electromagnetic field are equal to those for the scalar and vector gravitational perturbations respectively. Therefore we get the same results that in Ref.\ \cite{Natario:2004jd} for the AdS normal frequencies. Thus for the modes \textbf{II} we obtain the following values \cite{Natario:2004jd}
\begin{equation} \label{e: AdS normal frequencies}
\tilde{\omega} = (D + l -2 + 2n),  \qquad n =0,1,2,\dots,
\end{equation} 
for $D \geq 4$. While for the modes \textbf{I} we find that for $D \geq 6$ the AdS normal frequencies are equal to \cite{Natario:2004jd}
\begin{equation}
\tilde{\omega} = (D + l -3 + 2n).
\end{equation}
In $D=4$ its values are equal to \cite{Natario:2004jd}, \cite{Cardoso:2003cj}
\begin{equation} \label{e: AdS normal frequencies I}
\tilde{\omega} = (l + 2 + 2n),  
\end{equation}
and when $D=5$ we get that the spectrum of the AdS normal frequencies for modes \textbf{I} is continuous, as for scalar type gravitational perturbations, hence $\omega \in \mathbb{R}$ \cite{Natario:2004jd}.

The previous values for the normal modes $\omega$ are obtained by analyzing the master functions $\Phi_1(x)$ and $\Phi_2(x)$ of the Schr\"odinger type equations (\ref{e:Schrodinger-equation-I-II}). Also, we can impose the boundary conditions of the AdS modes on the radial functions $R_1$ and $R_2$. In the rest of this section we investigate this issue.

We can exactly solve the differential equations for the radial functions $R_1$ (Eq.\ (\ref{e:radial-I})) and $R_2$ (Eq.\ (\ref{e:radial-II})) as follows. In AdS spacetime these equations take the form
\begin{align} \label{e:radial AdS}
&(1+z^2)\frac{{\rm d}^2 R_1}{{\rm d} z^2} + \left(\frac{D}{z}(1+z^2) + 2z \right)\frac{{\rm d} R_1}{{\rm d} z}  \nonumber \\ 
&+ \left( \frac{D-2}{z^2}(1+z^2)  + 2(D-2) + \frac{\tilde{\omega}^2}{1+z^2} - \frac{l(l+D-3)}{z^2} \right)R_1=0, \nonumber \\
&(1+z^2)\frac{{\rm d}^2 R_2}{{\rm d} z^2} + \left(\frac{D -4}{z}(1+z^2) + 2z \right)\frac{{\rm d} R_2}{{\rm d} z} \\ 
& \qquad \qquad \qquad \quad + \left( \frac{\tilde{\omega}^2}{1+z^2} - \frac{(l+1)(l+D-4)}{z^2}\right)R_2=0.\nonumber
\end{align} 
Making the changes of variable $x=\arctan(z)$, $y=\sin^2 (x)$, and taking $R_1$ and $R_2$ as \cite{Natario:2004jd}, \cite{Cardoso:2003cj}
\begin{equation}
R_N = y^{C_N} (1-y)^{B_N} \tilde{R}_N, \qquad N=1,2, 
\end{equation} 
we get that the functions $\tilde{R}_N$ satisfy the hypergeometric differential equation \cite{b:DE-books}, \cite{b:special functions}
\begin{equation} \label{e:hypergeometric-differential}
y(1-y) \frac{{\rm d}^2 \tilde{R}_N}{{\rm d}y^2} + (c_N - (a_N +b_N + 1)y)\frac{{\rm d} \tilde{R}_N}{{\rm d}y} - a_N b_N  \tilde{R}_N  = 0,
\end{equation} 
where 
\begin{align} \label{e: a b c values}
&a_1 = C_1 + B_1 + \frac{\tilde{\omega}}{2}, \quad &a_2 = C_2 +  \frac{\tilde{\omega}}{2},\qquad\nonumber\\
&b_1 = C_1 + B_1 - \frac{\tilde{\omega}}{2}, \quad &b_2 = C_2 -  \frac{\tilde{\omega}}{2},\qquad\nonumber\\
&c_1 = 2 C_1 + \frac{D+1}{2},\,  &c_2 = 2 C_2 + \frac{D-3}{2},
\end{align}
with
\begin{eqnarray}
&B_1 = \left\{ \begin{array}{l} \frac{D-2}{2} , \\ \\ \frac{3}{2}, \end{array}\right. \qquad \qquad  &B_2 = 0, \\
&C_1 = \left\{ \begin{array}{l} \frac{l-1}{2} , \\ \\ - \frac{l+D-2}{2}, \end{array}\right. \qquad &C_2 = \left\{ \begin{array}{l} \frac{l+1}{2} , \\ \\ - \frac{l+D-4}{2}. \end{array}\right. \nonumber
\end{eqnarray} 
Depending on the values chosen for $B_N$, and $C_N$, the quantities $a_N$, $b_N$, and $c_N$ can take several values. Therefore we can write the two solutions to Eqs.\ (\ref{e:radial AdS}) in several equivalent forms. 

We first analyze the modes \textbf{II}, studying in detail the case $B_2=0$, and $C_2=(l+1)/2$. For these values of $B_2$, and $C_2$, the quantity $c_2$ is an integer (for odd $D$) or an half-integer (for even $D$). In both cases the first solution of Eq.\ (\ref{e:hypergeometric-differential}) is given by \cite{b:DE-books}
\begin{equation}
\tilde{R}_2^{(1)}={}_{2}F_{1}(a_2,b_2;c_2;y),
\end{equation} 
and the second  solution when $c_2$ is a half-integer is equal to
\begin{equation}\label{e:solution-two-1}
\tilde{R}_2^{(2)}=y^{1-c_2} {}_{2}F_{1}(a_2-c_2+1,b_2-c_2+1;2-c_2;y).
\end{equation} 
For $c_2$ an integer it takes the form
\begin{align} \label{e:solution-two-2}
 \tilde{R}_2^{(2)} &= {}_{2}F_{1}(a_2,b_2;c_2;y) \ln (y) \nonumber \\
& + \frac{(c_2 - 1)!}{\Gamma(a_2) \Gamma(b_2)} \sum_{s=1}^{c_2 - 1} (-1)^{s-1} (s - 1)! \frac{\Gamma(a_2 - s) \Gamma(b_2 -s)}{(c_2 -s - 1)!} y^{-s} \nonumber \\
&  + \sum_{s=0}^{\infty} \frac{(a_2)_s (b_2)_s}{s! (c_2)_s} y^s \left[\psi(a_2 + s) + \psi(b_2 + s) - \psi(c_2 + s) - \psi(1 + s) \right. \nonumber \\
& \left. \hspace{0cm} - \psi(a_2 -  c_2 +1) - \psi(b_2 - c_2 + 1) +\psi(1) + \psi(c_2-1)\right],
\end{align} 
where $\psi(z)={\rm d} \ln \Gamma(z)/{\rm d}z$, $(a)_0=1$ and $(a)_s=(a)(a+1)\cdots(a+s-1)$.

A straightforward analysis shows that the radial functions including $\tilde{R}_2^{(2)}$ (given in Eqs.\ (\ref{e:solution-two-1}) or (\ref{e:solution-two-2})) are divergent at the origin of the AdS spacetime due to presence of the factors $y^{1-l/2-(D-2)/2}$; while those including  $\tilde{R}_2^{(1)}$ are well behaved there. Thus, to satisfy the boundary condition (i) of the AdS normal modes, we must choose the solution regular at $y=0$ of the second differential equation in (\ref{e:radial AdS}); it is given by
\begin{equation}
R_2 = y^{\tfrac{l+1}{2}}  {}_{2}F_{1}(a_2,b_2;c_2;y).
\end{equation} 

It is easy show that $r \to \infty$ as $y \to 1$  because the relation between $r$ and $y$ is $r=Ly^{1/2}/(1-y)^{1/2}$. Also from Eq.\ (\ref{e: a b c values}) we find that $a_2$, $b_2$, and $c_2$ hold 
\begin{equation}
c_2-b_2-a_2 = \frac{D-3}{2} .
\end{equation}  
Thus, for $D \geq 4$, we find \cite{b:DE-books}, \cite{b:special functions}
\begin{equation}
\lim_{y \to 1} R_2 = \frac{\Gamma(c_2) \Gamma(c_2-b_2-a_2)}{\Gamma(c_2-a_2) \Gamma(c_2-b_2)}.
\end{equation} 
To satisfy the boundary condition $R_2(y=1) = 0$, we need to impose the conditions
\begin{equation}
c_2-a_2 = -n, \qquad \textrm{or} \qquad c_2-b_2=-n, 
\end{equation} 
and therefore from these we get for the AdS normal frequencies
\begin{equation} \label{e: AdS normal II}
\tilde{\omega} = (D + l -2 + 2n), \qquad \tilde{\omega} = - (D + l -2 + 2n).
\end{equation} 

In the following we study the frequencies with $\re (\omega)> 0 $; hence we only consider the first expression of the previous equation. These AdS normal frequencies are equal to those of Eq.\ (\ref{e: AdS normal frequencies}) which we calculated imposing the boundary condition on the master function $\Phi_1(x)$ as in Ref.\ \cite{Natario:2004jd}. In four dimensions they are also equal to those calculated in \cite{Cardoso:2003cj}. Moreover, the AdS normal frequencies (\ref{e: AdS normal II}) are obtained when we take $B_2=0$, and $C_2 = - (l + D - 4)/2$.

A similar procedure to that used for the modes \textbf{II} shows that for the modes \textbf{I}, when $B_1=3/2$ and $C_1=(l-1)/2$, the regular solution at $r=0$ is equal to
\begin{equation} \label{e: radial modes I}
R_1 = (1-y)^{\tfrac{3}{2}} y^{\tfrac{l-1}{2}} {}_{2}F_{1}(a_1,b_1;c_1;y).
\end{equation} 
From Eq.\ (\ref{e: a b c values}) we get that the quantities $a_1$, $b_1$, and $c_1$ satisfy 
\begin{equation}
c_1 -a_1 - b_1 = \frac{D-5}{2}.
\end{equation} 
Therefore, for $D\geq 6$, the hypergeometric function which appears in (\ref{e: radial modes I}) holds
\begin{equation}
\lim_{y \to 1} {}_{2}F_{1}(a_1,b_1;c_1;y) = \frac{\Gamma(c_1) \Gamma(c_1-b_1-a_1)}{\Gamma(c_1-a_1) \Gamma(c_1-b_1)}.
\end{equation}
Nevertheless, the radial function $R_1$ given by (\ref{e: radial modes I}) automatically satisfies the boundary condition at infinity of the AdS normal modes ($R_1(y=1)=0$), due to the factor $(1-y)^{3/2}$ and therefore there is no constraint on the frequencies  for $D \geq 6$. 

For $D=5$, exploiting the property of the hypergeometric function \cite{b:DE-books}, \cite{b:special functions}
\begin{align}
{}_{2}F_{1}(a,b;c;y) = \frac{\Gamma(a+b)}{\Gamma(a) \Gamma(b)} \sum_{k=0}^{\infty} \frac{(a)_k (b)_k}{(k!)^2} & \left[ 2 \psi(k+1) - \psi(a+k)  \right.  \\ 
\qquad & \left. -\psi(b+k)  - \ln (1-y) \right] (1-y)^k , \nonumber
\end{align} 
which is valid when $c-a-b=0$, we can show that the radial function $R_1$ regular at $r=0$ satisfies $R_1(y=1)=0$. 

For $D=4$, from the property of the hypergeometric function \cite{b:DE-books}, \cite{b:special functions}
\begin{equation}
{}_{2}F_{1}(a,b;c;y) = (1-y)^{c-a-b} {}_{2}F_{1}(c-a,c-b;c;y),
\end{equation} 
we find that the radial function $R_1$ regular at $r=0$ takes the form
\begin{equation}
R_1 = (1-y)\, y^{\tfrac{l-1}{2}}\, {}_{2}F_{1}\left(\frac{l+1}{2} + \frac{\tilde{\omega}}{{2}},\frac{l+1}{2} - \frac{\tilde{\omega}}{{2}}; l + \frac{3}{2};y\right).
\end{equation}  
This radial function automatically fulfills the boundary condition as $r \to \infty$. Therefore in four and five dimensions, there is no constraint on the frequency of the field to satisfy the boundary condition at $y=1$.

The previous conclusions imply that for the modes \textbf{I} the spectrum of the AdS normal modes is continuous, thus $\omega \in \mathbb{R}$ when $D\geq 4$. This result is different from that obtained taking into account the findings of Nat\'ario and Schiappa \cite{Natario:2004jd}, (a continuous spectrum for the AdS normal frequencies of the modes \textbf{I} only for the 5-dimensional background).

To understand this fact we notice that the factor $1/r^{D/2}$ ($1/r^{(D-4)/2}$) in Eq.\ (\ref{e: R1 to Phi}) which relates $R_1$ ($R_2$) with $\Phi_1(x)$ ($\Phi_2(x)$) implies that the radial function $R_1$ ($R_2$) is more singular than $\Phi_1(x)$ ($\Phi_2(x)$) at $y=0$. This factor also produces that the radial function $R_1$ ($R_2$) is less singular than $\Phi_1(x)$ ($\Phi_2(x)$) at $y=1$.

For example, in the 4-dimensional AdS spacetime, the effective potentials for modes \textbf{I} and \textbf{II} are equal \cite{Ishibashi:2004wx}; so $\Phi_1(x)=\Phi_2(x)=\Phi(x)$ and from Eqs.\ (\ref{e: R1 to Phi}) we obtain
\begin{align}
R_1 = \frac{\Phi(x)}{r^2}, \qquad
R_2  = \Phi(x).
\end{align}
Taking into account the relation between $r$ and $y$ we get
\begin{equation} \label{e: relation R1 R2}
R_1 = \frac{1-y}{L^2 y} R_2,
\end{equation} 
where
\begin{equation}
R_2 = y^{\tfrac{l+1}{2}} {}_{2}F_{1}\left(\frac{l+1}{2} + \frac{\tilde{\omega}}{{2}},\frac{l+1}{2} - \frac{\tilde{\omega}}{{2}}; l + \frac{3}{2};y\right).
\end{equation} 

The factor $(1-y)$ in Eq.\ (\ref{e: relation R1 R2}) implies that the function $R_1$ satisfies the boundary condition at spatial infinity without imposing a constraint on the frequency of the field, while the radial function $R_2$ fulfills it, if and only if we impose a condition on the frequency of the field. Thus the result obtained for the AdS normal frequencies of the modes \textbf{I} depends upon if we impose the boundary conditions on the radial function $R_1$ or the master function $\Phi_1(x)$. We believe that this point must be studied further.

A similar analysis for the gravitational perturbations is more complicated due to the complex transformations used to simplify the equations of motion to Schr\"o\-ding\-er type equations \cite{Kodama:2003jz}. 

\begin{table}[t]
\caption{Values of $\mathcal{A}$ and $\mathcal{B}$. Here Grav. stands for \textit{gravitational perturbation} and em stands for \textit{electromagnetic field}. The symbol $m$ stands for the mass of the Klein Gordon field and $\tilde{m} = m L$.} 
\label{t:Table1}
\centering
\begin{tabular}{lll} 
\hline\noalign{\smallskip} \noalign{\smallskip}
Field & $\mathcal{A}$ & $\mathcal{B}$ \\ \hline \noalign{\smallskip}
Massive Klein Gordon & $\tilde{m}\,{}^{2} + \tfrac{D(D-2)}{4}$ & $l(l+D-3) + \tfrac{(D-2)(D-4)}{4}$ \\  \noalign{\smallskip} \hline \noalign{\smallskip}
Grav. tensor type & $\tfrac{D(D-2)}{4}$ & $l(l+D-3) + \tfrac{(D-2)(D-4)}{4}$ \\  \noalign{\smallskip} \hline \noalign{\smallskip}
Grav. vector type and em modes \textbf{II} & $\tfrac{(D-2)(D-4)}{4}$ & $l(l+D-3) + \tfrac{(D-2)(D-4)}{4}$ \\  \noalign{\smallskip} \hline \noalign{\smallskip}
Grav. scalar type and em modes \textbf{I} & $\tfrac{(D-4)(D-6)}{4}$ & $l(l+D-3) + \tfrac{(D-2)(D-4)}{4}$ \\ 
\noalign{\smallskip}\hline
\end{tabular} 
\end{table}

To finish this section we note that the effective potentials for the electromagnetic, gravitational and massive Klein-Gordon perturbations moving in AdS spacetime can be written in the form 
\begin{equation}
V(x) = \frac{1}{L^2} \left[ \frac{\mathcal{A}}{\cos^2(x/L)} + \frac{\mathcal{B}}{\sin^2(x/L)} \right],
\end{equation} 
where the quantities $\mathcal{A}$ and $\mathcal{B}$ for these fields are presented in Table \ref{t:Table1}. Thus the effective potentials are of P\"oschl-Teller type \cite{b:Poschl Teller potential}. In Table 1 of Ref.\ \cite{Lopez-Ortega:2006} another list of spacetimes for which their effective potentials take the P\"oschl-Teller form is given.

\section{Discussion}
\label{Section 7}

Owing to the stability against gravitational perturbations of the $D$-dimensional SdS and SadS is not completely proven, our results hold if and only if the black hole in consideration is stable. We believe that the stability of the higher dimensional backgrounds analyzed in this paper must be studied further.

In this framework, we should produce numerical calculations that prove (or disprove) our analytical findings. In particular, just as was done in \cite{Yoshida:2003zz}, \cite{Konoplya:2004uk} for the asymptotic QN frequencies of the 4-dimensional black holes found in Refs.\ \cite{Natario:2004jd}, \cite{Cardoso:2004up}, a numerical computation must be performed to verify the corresponding frequencies of an electromagnetic field moving in 6-dimensional Schwarzschild and SdS backgrounds, (they are shown in Eqs.\ (\ref{e: S QN frequencies 6D}) and (\ref{e:QN frequency 6D SdS}) respectively). We also believe that the case of the 5-di\-men\-sion\-al Schwarzschild black hole must be studied using other analytical methods to explain the numerical results of \cite{Cardoso:2003vt}.

For the 5-dimensional SadS black hole the asymptotic QN frequencies of the modes \textbf{I} have a continuous spectrum as the scalar gravitational perturbations \cite{Natario:2004jd}. To investigate in more detail their properties we must calculate numerically the QNMs of the gravitational and electromagnetic perturbations for this background. In particular, an interesting problem is to verify if the spectrum of the QN frequencies is continuous when $n \to \infty$.

Since the parameters $j_1$ and $j_2$ of the electromagnetic field depend upon the spacetime dimension $D$, from expressions obtained for its asymptotic QN frequencies, we find that they also show a similar dependence, while for the gravitational perturbations moving in $D$-di\-men\-sion\-al uncharged black holes the parameters $j$ and therefore the real part of their asymptotic QN frequencies do not depend on the spacetime dimension \cite{Motl:2003cd}, \cite{Birmingham:2003rf}, \cite{Natario:2004jd}.

Notice that the final expressions for the asymptotic QN frequencies of an gravitational perturbation propagating in Schwarzschild, SdS, and SadS black holes given in Ref.\ \cite{Natario:2004jd} hold for the cases $j=0$ and $j=2$. The expressions for the asymptotic QN frequencies already obtained in this paper are valid for more values of $j$ and for the SdS black hole we can obtain the results of Nat\'ario and Schiappa \cite{Natario:2004jd} as a limit of our Eqs.\ (\ref{e: QNMs SdS 6}) and (\ref{e: QNMs SdS 5D}).

As already noted in \cite{Natario:2004jd}, the de Sitter limit of the asymptotic QN frequencies for the SdS black hole must be carefully taken, since we do not always get the behavior analytically found in Refs.\ \cite{Natario:2004jd}, \cite{Lopez-Ortega:2006}, \cite{Lopez-Ortega:2006ig} for the frequencies of the de Sitter background. Furthermore, our findings after taking this limit are different from those of Nat\'ario and Schiappa even for the gravitational perturbations.

An extension of our work is to study the asymptotic QN frequencies of the gravitational and electromagnetic perturbations in other spacetimes which are as\-ymp\-tot\-ical\-ly flat, de Sitter and anti-de Sitter as in Refs.\ \cite{Ghosh:2005aq}, \cite{Das:2004db}. Another extension of our results is to calculate the first order corrections to the QN frequencies of an electromagnetic field moving in $D$-dimensional SdS and SadS black holes as in Refs.\ \cite{Cardoso:2003vt}, \cite{Musiri:2005ev} was done for the electromagnetic perturbations of the Schwarzschild background and the gravitational perturbations of the SadS spacetime respectively.

In AdS spacetime we should investigate if the boundary conditions of the AdS normal modes must be imposed on the radial function of the fields or the master variable which appear in the Schr\"odinger type equation, because we get different results for them that depend on this choice, as shown in Section \ref{Section 6}. Furthermore, instead of the boundary condition (ii) used in previous section, we may impose: \textit{the flux of the field is zero at spatial infinity} \cite{Birmingham:2001pj}, \cite{Dasgupta:1998jg}. Do we get different results for the AdS normal frequencies? This question must be studied further.

\section{Acknowledgements}

I thank Dr.\ M.\ A.\ P\'erez Ang\'on for his interest in this paper and also for proofreading the manuscript. I also thank Referees for their comments. This work was supported by CONACyT and SNI (M\'exico).






\begin{thebibliography}{99}



\bibitem{Frolov:2002xf}
  V.~P.~Frolov and D.~Stojkovic,
  Phys.\ Rev.\ D {\bf 67}, 084004 (2003)
  [arXiv:gr-qc/0211055];
  V.~P.~Frolov and D.~Stojkovic,
  Phys.\ Rev.\ D {\bf 68}, 064011 (2003)
  [arXiv:gr-qc/0301016].

\bibitem{Kodama:2003jz}
  H.~Kodama and A.~Ishibashi,
  Prog.\ Theor.\ Phys.\  {\bf 110}, 701 (2003)
  [arXiv:hep-th/0305147];
  H.~Kodama, A.~Ishibashi and O.~Seto,
  Phys.\ Rev.\ D {\bf 62}, 064022 (2000)
  [arXiv:hep-th/0004160].

\bibitem{Kodama:2003kk}
  H.~Kodama and A.~Ishibashi,
  Prog.\ Theor.\ Phys.\  {\bf 111}, 29 (2004)
  [arXiv:hep-th/0308128].

\bibitem{Crispino:2000jx}
  L.~C.~B.~Crispino, A.~Higuchi and G.~E.~A.~Matsas,
  Phys.\ Rev.\ D {\bf 63}, 124008 (2001)
  [arXiv:gr-qc/0011070].

\bibitem{Lopez-Ortega:2003gi}
  A.~Lopez-Ortega,
  Gen.\ Rel.\ Grav.\  {\bf 35}, 1785 (2003).

\bibitem{Gibbons:2002pq}
  G.~Gibbons and S.~A.~Hartnoll,
  Phys.\ Rev.\ D {\bf 66}, 064024 (2002)
  [arXiv:hep-th/0206202];
  I.~P.~Neupane, 
  Phys.\ Rev.\ D {\bf 69}, 084011 (2004)
  [arXiv:hep-th/0302132].

\bibitem{Ishibashi:2003ap}
  A.~Ishibashi, and H.~Kodama,
  Prog.\ Theor.\ Phys.\  {\bf 110}, 901 (2003)
  [arXiv:hep-th/0305185].

\bibitem{Gleiser:2005ra}
R.~J.~Gleiser and G.~Dotti,
  Phys.\ Rev.\ D {\bf 72}, 124002 (2005)
  [arXiv:gr-qc/0510069];
  G.~Dotti, and R.~J.~Gleiser, 
  Class.\ Quant.\ Grav.\  {\bf 22} (2005) L1
  [arXiv:gr-qc/0409005];
  G.~Dotti, and R.~J.~Gleiser, 
  Phys.\ Rev.\ D {\bf 72}, 044018 (2005)
  [arXiv:gr-qc/0503117].

\bibitem{Cardoso:2005vb}
  V.~Cardoso, M.~Cavaglia and L.~Gualtieri,
  Phys.\ Rev.\ Lett.\  {\bf 96}, 071301 (2006)
  [arXiv:hep-th/0512002];
  V.~Cardoso, M.~Cavaglia and L.~Gualtieri,
  JHEP {\bf 0602}, 021 (2006)
  [arXiv:hep-th/0512116].

\bibitem{Cornell:2005ux}
  A.~S.~Cornell, W.~Naylor and M.~Sasaki,
  JHEP {\bf 0602}, 012 (2006)
  [arXiv:hep-th/0510009].

\bibitem{Creek:2006ia}
  S.~Creek, O.~Efthimiou, P.~Kanti and K.~Tamvakis,
  Phys.\ Lett.\ B {\bf 635}, 39 (2006)
  [arXiv:hep-th/0601126].

\bibitem{Motl:2003cd}
  L.~Motl and A.~Neitzke,
  Adv.\ Theor.\ Math.\ Phys.\  {\bf 7}, 307 (2003)
  [arXiv:hep-th/0301173].

\bibitem{Birmingham:2003rf}
  D.~Birmingham,
  Phys.\ Lett.\ B {\bf 569}, 199 (2003)
  [arXiv:hep-th/0306004].

\bibitem{Natario:2004jd}
  J.~Natario and R.~Schiappa,
  Adv.\ Theor.\ Math.\ Phys.\  {\bf 8}, 1001 (2004)
  [arXiv:hep-th/0411267].

\bibitem{Du:2004jt}
  D.~P.~Du, B.~Wang and R.~K.~Su,
  Phys.\ Rev.\ D {\bf 70}, 064024 (2004)
  [arXiv:hep-th/0404047].

\bibitem{Aros:2002te}
  R.~Aros, C.~Martinez, R.~Troncoso and J.~Zanelli,
  Phys.\ Rev.\ D {\bf 67}, 044014 (2003)
  [arXiv:hep-th/0211024].

\bibitem{Molina:2003ff}
  C.~Molina,
  Phys.\ Rev.\ D {\bf 68}, 064007 (2003)
  [arXiv:gr-qc/0304053].

\bibitem{Ida:2002zk}
  D.~Ida, Y.~Uchida and Y.~Morisawa,
  Phys.\ Rev.\ D {\bf 67}, 084019 (2003)
  [arXiv:gr-qc/0212035];
  Y.~Morisawa and D.~Ida,
  Phys.\ Rev.\ D {\bf 71}, 044022 (2005)
  [arXiv:gr-qc/0412070].

\bibitem{Konoplya:2003ii}
  R.~A.~Konoplya,
  Phys.\ Rev.\ D {\bf 68}, 024018 (2003)
  [arXiv:gr-qc/0303052];
  R.~A.~Konoplya,
  Phys.\ Rev.\ D {\bf 68}, 124017 (2003)
  [arXiv:hep-th/0309030];
  P.~Kanti and R.~A.~Konoplya,
  Phys.\ Rev.\ D {\bf 73}, 044002 (2006)
  [arXiv:hep-th/0512257];
  P.~Kanti, R.~A.~Konoplya and A.~Zhidenko,
  arXiv:gr-qc/0607048.


\bibitem{Ghosh:2005aq}
  A.~Ghosh, S.~Shankaranarayanan and S.~Das,
  Class.\ Quant.\ Grav.\  {\bf 23}, 1851 (2006)
  [arXiv:hep-th/0510186].

\bibitem{Cardoso:2003vt}
  V.~Cardoso, J.~P.~S.~Lemos and S.~Yoshida,
  Phys.\ Rev.\ D {\bf 69}, 044004 (2004)
  [arXiv:gr-qc/0309112].

\bibitem{Lopez-Ortega:2006} 
  A.~L\'opez-Ortega, 
  arXiv:gr-qc/0605027. To be published in GRG.

\bibitem{Kokkotas:1999bd}
  K.~D.~Kokkotas and B.~G.~Schmidt,
  Living Rev.\ Rel.\  {\bf 2}, 2 (1999)
  [arXiv:gr-qc/9909058];
H.~P.~Nollert, 
Class.\ Quantum Grav.\ {\bf 16}, R159 (1999); 
  E.~Berti,
  arXiv:gr-qc/0411025.

\bibitem{Birmingham:2001pj}
  D.~Birmingham, I.~Sachs and S.~N.~Solodukhin,
  Phys.\ Rev.\ Lett.\  {\bf 88}, 151301 (2002)
  [arXiv:hep-th/0112055];
  D.~Birmingham,
  Phys.\ Rev.\ D {\bf 64}, 064024 (2001)
  [arXiv:hep-th/0101194].

\bibitem{Dasgupta:1998jg}   
  A.~Dasgupta,
  Phys.\ Lett.\ B {\bf 445}, 279 (1999)
  [arXiv:hep-th/9808086];
  S.~Das and A.~Dasgupta,
  JHEP {\bf 9910}, 025 (1999)
  [arXiv:hep-th/9907116].

\bibitem{Horowitz:1999jd}
  G.~T.~Horowitz and V.~E.~Hubeny,
  Phys.\ Rev.\ D {\bf 62}, 024027 (2000)
  [arXiv:hep-th/9909056].

\bibitem{Cardoso:2001bb}
  V.~Cardoso and J.~P.~S.~Lemos,
  Phys.\ Rev.\ D {\bf 64}, 084017 (2001)
  [arXiv:gr-qc/0105103].

\bibitem{Siopsis:2004up}
  G.~Siopsis,
  Phys.\ Lett.\ B {\bf 590}, 105 (2004)
  [arXiv:hep-th/0402083];
  G.~Siopsis,
  Nucl.\ Phys.\ B {\bf 715}, 483 (2005)
  [arXiv:hep-th/0407157].

\bibitem{Abdalla:2002hg}
  E.~Abdalla, K.~H.~C.~Castello-Branco and A.~Lima-Santos,
  Phys.\ Rev.\ D {\bf 66}, 104018 (2002)
  [arXiv:hep-th/0208065].

\bibitem{Abdalla:2002rm}
  E.~Abdalla, B.~Wang, A.~Lima-Santos and W.~G.~Qiu,
  Phys.\ Lett.\ B {\bf 538}, 435 (2002)
  [arXiv:hep-th/0204030].

\bibitem{Hod:1998vk}
  S.~Hod,
  Phys.\ Rev.\ Lett.\  {\bf 81}, 4293 (1998)
  [arXiv:gr-qc/9812002].

\bibitem{Dreyer:2002vy}
  O.~Dreyer,
  Phys.\ Rev.\ Lett.\  {\bf 90}, 081301 (2003)
  [arXiv:gr-qc/0211076].

\bibitem{Birmingham:2003wa}
  D.~Birmingham, S.~Carlip and Y.~j.~Chen,
  Class.\ Quant.\ Grav.\  {\bf 20}, L239 (2003)
  [arXiv:hep-th/0305113];
  G.~Kunstatter,
  Phys.\ Rev.\ Lett.\  {\bf 90}, 161301 (2003)
  [arXiv:gr-qc/0212014].

\bibitem{Brady:1999wd}
  P.~R.~Brady, C.~M.~Chambers, W.~G.~Laarakkers and E.~Poisson,
  Phys.\ Rev.\ D {\bf 60}, 064003 (1999)
  [arXiv:gr-qc/9902010];
  T.~R.~Choudhury and T.~Padmanabhan,
  Phys.\ Rev.\ D {\bf 69}, 064033 (2004)
  [arXiv:gr-qc/0311064].

\bibitem{Lepe:2004kv}
  S.~Lepe and J.~Saavedra,
  Phys.\ Lett.\ B {\bf 617}, 174 (2005)
  [arXiv:gr-qc/0410074];
  J.~Saavedra,
  arXiv:gr-qc/0508040.

\bibitem{Lopez-Ortega:2005ep}
  A.~Lopez-Ortega,
  Gen.\ Rel.\ Grav.\  {\bf 37}, 167 (2005);
  S.~Fernando,
  Gen.\ Rel.\ Grav.\  {\bf 36}, 71 (2004)
  [arXiv:hep-th/0306214].

\bibitem{Vanzo:2004fy}
  L.~Vanzo and S.~Zerbini,
  Phys.\ Rev.\ D {\bf 70}, 044030 (2004)
  [arXiv:hep-th/0402103].

\bibitem{Cardoso:2001hn}
  V.~Cardoso and J.~P.~S.~Lemos,
  Phys.\ Rev.\ D {\bf 63}, 124015 (2001)
  [arXiv:gr-qc/0101052];
  V.~Cardoso and J.~P.~S.~Lemos,
  Phys.\ Rev.\ D {\bf 67}, 084020 (2003)
  [arXiv:gr-qc/0301078].

\bibitem{Schutz:1985yo}  
B.~F.~Schutz, and C.~M.~Will,
Astrophys.\ J.\ 291, L33 (1985).

\bibitem{Iyer:1986np}
  S.~Iyer and C.~M.~Will,
  Phys.\ Rev.\ D {\bf 35}, 3621 (1987).

\bibitem{Iyer:1986nq}
  S.~Iyer,
  Phys.\ Rev.\ D {\bf 35}, 3632 (1987).
  K.~D.~Kokkotas and B.~F.~Schutz,
  Phys.\ Rev.\ D {\bf 37}, 3378 (1988);
  E.~Seidel and S.~Iyer,
  Phys.\ Rev.\ D {\bf 41}, 374 (1990).

\bibitem{Andersson:1992yo}
N.~Andersson and S.~Linn\ae us,
Phys.\ Rev.\ D {\bf 46}, 4179 (1992);
  N.~Froeman, P.~O.~Froeman, N.~Andersson and A.~Hoekback,
  Phys.\ Rev.\ D {\bf 45}, 2609 (1992);
N.~Andersson, M.~E.~Araujo and B.~F.~Schutz,
Class.\ Quantum Grav.\ {\bf 10}, 735 (1993);
N.~Andersson, M.~E.~Araujo, and B.~F.~Schutz,
Class.\ Quantum Grav.\ {\bf 10}, 757 (1993).

\bibitem{Andersson PRSLA}
N.~Andersson,
Proc.\ Roy.\ Soc.\ Lond.\ A {\bf 439}, 47 (1992);
N.~Andersson,
Proc.\ Roy.\ Soc.\ Lond.\ A {\bf 442}, 427 (1993).

\bibitem{Cho:2005yc}
  H.~T.~Cho,
  Phys.\ Rev.\ D {\bf 73}, 024019 (2006)
  [arXiv:gr-qc/0512052];
  R.~G.~Daghigh and G.~Kunstatter,
  Class.\ Quant.\ Grav.\  {\bf 22}, 4113 (2005)
  [arXiv:gr-qc/0505044];
  R.~G.~Daghigh, G.~Kunstatter, D.~Ostapchuk and V.~Bagnulo,
  arXiv:gr-qc/0604073.

\bibitem{Andersson:2003fh}
  N.~Andersson and C.~J.~Howls,
  Class.\ Quant.\ Grav.\  {\bf 21}, 1623 (2004)
  [arXiv:gr-qc/0307020].

\bibitem{Mashhoon:1984yo}
V.~Ferrari and B.~Mashhoon
Phys.\ Rev.\ D {\bf 30}, 295 (1984);
%
V.~Ferrari and B.~Mashhoon
Phys.\ Rev.\ Lett.\ {\bf 52}, 1361 (1984);
%
H.~J.~Blome, and B.~Mashhoon, 
Phys.\ Lett.\ A {\bf 100}, 231, (1984).

\bibitem{Leaver:1985ax}
  E.~W.~Leaver,
  Proc.\ Roy.\ Soc.\ Lond.\ A {\bf 402}, 285 (1985);
E.~W.~Leaver,
Phys.\ Rev.\ D {\bf 41}, 2986 (1990).

\bibitem{Nollert:1993aa} 
H.~P.~Nollert, 
Phys.\ Rev.\ D {\bf 47}, 5253 (1993).

\bibitem{Onozawa:1995vu}
  H.~Onozawa, T.~Mishima, T.~Okamura and H.~Ishihara,
  Phys.\ Rev.\ D {\bf 53}, 7033 (1996)
  [arXiv:gr-qc/9603021].

\bibitem{Motl:2002hd}
  L.~Motl,
  Adv.\ Theor.\ Math.\ Phys.\  {\bf 6}, 1135 (2003)
  [arXiv:gr-qc/0212096].

\bibitem{Cardoso:2004up}
  V.~Cardoso, J.~Natario and R.~Schiappa,
  J.\ Math.\ Phys.\  {\bf 45}, 4698 (2004)
  [arXiv:hep-th/0403132].

\bibitem{Das:2004db}
  S.~Das and S.~Shankaranarayanan,
  Class.\ Quant.\ Grav.\  {\bf 22}, L7 (2005)
  [arXiv:hep-th/0410209].

\bibitem{Kettner:2004aw}
  J.~Kettner, G.~Kunstatter and A.~J.~M.~Medved,
  Class.\ Quant.\ Grav.\  {\bf 21}, 5317 (2004)
  [arXiv:gr-qc/0408042].

\bibitem{Krasnov:2004ki}
  K.~Krasnov and S.~N.~Solodukhin,
  Adv.\ Theor.\ Math.\ Phys.\  {\bf 8}, 421 (2004)
  [arXiv:hep-th/0403046].

\bibitem{Castello-Branco:2003jp}
  K.~H.~C.~Castello-Branco and E.~Abdalla,
  arXiv:gr-qc/0309090.

\bibitem{Chen:2005pv}
  S.~B.~Chen and J.~L.~Jing,
  Class.\ Quant.\ Grav.\  {\bf 22}, 2159 (2005)
  [arXiv:gr-qc/0511106];
  S.~b.~Chen and J.~l.~Jing,
  Class.\ Quant.\ Grav.\  {\bf 22}, 533 (2005)
  [arXiv:gr-qc/0409013].

\bibitem{Park:2005rw}
  D.~K.~Park,
  Phys.\ Lett.\ B {\bf 633}, 613 (2006)
  [arXiv:hep-th/0511159].

\bibitem{Chakrabarti:2005cm}
  S.~K.~Chakrabarti and K.~S.~Gupta,
  arXiv:hep-th/0506133.

\bibitem{Yoshida:2003zz}
  S.~Yoshida and T.~Futamase,
  Phys.\ Rev.\ D {\bf 69}, 064025 (2004)
  [arXiv:gr-qc/0308077].

\bibitem{Konoplya:2004uk}
  R.~A.~Konoplya and A.~Zhidenko,
  JHEP {\bf 0406}, 037 (2004)
  [arXiv:hep-th/0402080];
  A.~Zhidenko,
  Class.\ Quant.\ Grav.\  {\bf 21}, 273 (2004)
  [arXiv:gr-qc/0307012].

\bibitem{Cardoso:2003cj}
  V.~Cardoso, R.~Konoplya and J.~P.~S.~Lemos,
  Phys.\ Rev.\ D {\bf 68}, 044024 (2003)
  [arXiv:gr-qc/0305037].

\bibitem{Berti:2003ud}
  E.~Berti and K.~D.~Kokkotas,
  Phys.\ Rev.\ D {\bf 67}, 064020 (2003)
  [arXiv:gr-qc/0301052].

\bibitem{Khriplovich:2005wf}
  I.~B.~Khriplovich and G.~Y.~Ruban,
  arXiv:gr-qc/0511056;
  K.~H.~C.~Castello-Branco, R.~A.~Konoplya and A.~Zhidenko,
  Phys.\ Rev.\ D {\bf 71}, 047502 (2005)
  [arXiv:hep-th/0411055];
  J.~l.~Jing,
  Phys.\ Rev.\ D {\bf 71}, 124006 (2005)
  [arXiv:gr-qc/0502023].

\bibitem{Musiri:2005ev}
  S.~Musiri, S.~Ness and G.~Siopsis,
  Phys.\ Rev.\ D {\bf 73}, 064001 (2006)
  [arXiv:hep-th/0511113].

\bibitem{Musiri:2003bv}
  S.~Musiri and G.~Siopsis,
  Class.\ Quant.\ Grav.\  {\bf 20}, L285 (2003)
  [arXiv:hep-th/0308168];
  S.~Musiri and G.~Siopsis,
  Phys.\ Lett.\ B {\bf 576}, 309 (2003)
  [arXiv:hep-th/0308196];
  G.~Siopsis,
  Phys.\ Lett.\ B {\bf 590}, 105 (2004)
  [arXiv:hep-th/0402083].

\bibitem{Domagala:2004jt}
  M.~Domagala and J.~Lewandowski,
  Class.\ Quant.\ Grav.\  {\bf 21}, 5233 (2004)
  [arXiv:gr-qc/0407051];
  K.~A.~Meissner,
  Class.\ Quant.\ Grav.\  {\bf 21}, 5245 (2004)
  [arXiv:gr-qc/0407052];
  O.~Dreyer, F.~Markopoulou and L.~Smolin,
  arXiv:hep-th/0409056;
  S.~Alexandrov,
  arXiv:gr-qc/0408033;
  I.~B.~Khriplovich,
  Int.\ J.\ Mod.\ Phys.\ D {\bf 14}, 181 (2005)
  [arXiv:gr-qc/0407111].

\bibitem{Breitenlohner:1982bm}
  P.~Breitenlohner and D.~Z.~Freedman,
  Phys.\ Lett.\ B {\bf 115}, 197 (1982);
  P.~Breitenlohner and D.~Z.~Freedman,
  Annals Phys.\  {\bf 144}, 249 (1982).

\bibitem{Burgess:1984ti}
  C.~P.~Burgess and C.~A.~Lutken,
  Phys.\ Lett.\ B {\bf 153}, 137 (1985);
  L.~Mezincescu and P.~K.~Townsend,
  Annals Phys.\  {\bf 160}, 406 (1985).

\bibitem{Cotaescu:1998ts}
  I.~I.~Cot\u aescu,
  Mod.\ Phys.\ Lett.\ A {\bf 13}, 2923 (1998)
  [arXiv:gr-qc/9803042];
  I.~I.~Cot\u aescu,
  Phys.\ Rev.\ D {\bf 60}, 107504 (1999);
  I.~I.~Cot\u aescu,
  Phys.\ Rev.\ D {\bf 60}, 124006 (1999)
  [arXiv:gr-qc/9910004];
  I.~I.~Cot\u aescu,
  Int.\ J.\ Mod.\ Phys.\ A {\bf 19}, 2217 (2004)
  [arXiv:gr-qc/0306127].

\bibitem{Avis:1977yn}
  S.~J.~Avis, C.~J.~Isham and D.~Storey,
  Phys.\ Rev.\ D {\bf 18}, 3565 (1978).

\bibitem{Ishibashi:2004wx}
  A.~Ishibashi and R.~M.~Wald,
  Class.\ Quant.\ Grav.\  {\bf 21}, 2981 (2004)
  [arXiv:hep-th/0402184].


\bibitem{Guven hertz potential} R.~Guven, Class.\ Quantum Grav.\ {\bf 6}, 1961 (1989).





\bibitem{b:DE-books}  M.~Abramowitz and I.~A.~Stegun, {\it Handbook of Mathematical Functions, Graphs, and Mathematical Table}, Dover Publications, New York (1965). 


\bibitem{Lopez-Ortega:2006ig}
  A.~Lopez-Ortega,
  Gen.\ Rel.\ Grav.\  {\bf 38}, 743 (2006)
  [arXiv:gr-qc/0605022].


\bibitem{b:special functions} Z.~X.~Wang and  D.~R.~Guo, {\it Special Functions}, World Scientific Publishing, Singapore (1989);  N.~N.~Lebedev, {\it Special Functions and their Applications}, Dover Publications, New York (1972).


\bibitem{b:Poschl Teller potential}  G.~P\"oschl, and E.~Teller, Z.\ Phys.\ \textbf{83}, 143 (1933).


\end{thebibliography}
\end{document}